\documentclass[twocolumn,aps,prb]{revtex4-1}

\usepackage{graphicx}
\usepackage{amsmath}
\usepackage{url}
\usepackage{float}
\usepackage[english]{babel}
\usepackage{chngcntr}
\usepackage{natbib}
\usepackage{hyperref}
\usepackage{tikz}
\usepackage{xcolor}

\hypersetup{
    colorlinks,
    citecolor=blue,
    filecolor=blue,
    linkcolor=blue,
    urlcolor=blue
}

\begin{document}

\title{Interplay of spin-orbit and hyperfine interactions in dynamical nuclear polarization in semiconductor quantum dots}

\author{Marko J. Ran\v{c}i\'{c}}
\author{Guido Burkard}
\affiliation{Department of Physics, University of Konstanz, D-78457 Konstanz, Germany}

\date{\today}

\begin{abstract}
We theoretically study the interplay of spin-orbit and hyperfine interactions in dynamical nuclear polarization in two-electron semiconductor double quantum dots near the singlet $(S)$ - triplet $(T_+)$ anticrossing.
The goal of the scheme under study is to extend the singlet $(S)$ - triplet $(T_0)$ qubit decoherence time $T_2^{*}$ by dynamically transferring the polarization from the electron spins to the nuclear spins. 
This polarization transfer is achieved by cycling the electron spins over the $S-T_+$ anticrossing. Here, we investigate, both quantitatively and qualitatively, 
how this hyperfine mediated dynamical polarization transfer is influenced by the Rashba and Dresselhaus spin-orbit interaction. In addition to $T_2^*$, we determine the singlet return probability $P_s$, a quantity that can be measured in experiments. 
Our results suggest that the spin-orbit interaction establishes a mechanism that can polarize the nuclear spins in the opposite direction compared to hyperfine mediated nuclear spin polarization. 
In materials with relatively strong spin-orbit coupling, this interplay of spin-orbit and hyperfine mediated nuclear spin polarizations prevents any notable increase of the $S-T_0$ qubit decoherence time $T_2^{*}$.
\end{abstract}

\maketitle

\section{INTRODUCTION}

Electron spins in semiconductor quantum dots are considered to be excellent candidates for qubits [\onlinecite{a1}]. In order for a full scale quantum computer to be produced, a successful fulfillment of the DiVincenzo criteria [\onlinecite{a1b}] is necessary. 
Accurate qubit manipulation [\onlinecite{b2}, \onlinecite{Nowack}] and reliable state preparation [\onlinecite{1}] are some of the requirements that have been satisfied in the past years. Techniques for qubit identification and fast readout are also known, e.g., 
the spin readout for a two-electron double quantum dot is most commonly done in the regime of Pauli spin blockade [\onlinecite{a2}] using spin to charge conversion measurements [\onlinecite{a3}]. Still, one challenge remains - sufficiently isolating the 
qubit from the corruptive effects of its surroundings. 

Due to the influence of its surroundings, a qubit will irreversibly lose information. Different types of information losses happen on different time scales.
The time in which a qubit relaxes to a state of thermal equilibrium is the relaxation time $T_{1}$, whereas the time in which a qubit loses coherence due to the collective effects of its surroundings is the decoherence time $T_2^*$.
Although experimental and theoretical solutions for overcoming these information losses have been steadily developed for years, [\onlinecite{t18}-\onlinecite{i9}] 
overcoming qubit decoherence caused by a fluctuating nuclear spin bath is still an ongoing task.

Silicon [\onlinecite{kane}] and graphene [\onlinecite{Bjorn}] have stable isotopes with a zero nuclear spin. Therefore, they can be isotopically purified leaving only spin zero nuclei which do not contribute to the electron spin qubit decoherence. 
On the other hand, III-IV semiconductors, and particularly In$_x$Ga$_{1-x}$As structures only have stable isotopes with a non-zero nuclear spin. An electron confined in a typical In$_x$Ga$_{1-x}$As quantum dot interacts with $10^4-10^6$ nuclear 
spins, which contribute strongly to electron spin qubit decoherence. Optically [\onlinecite{Cekovic}-\onlinecite{Imamoglu2}] or electrically polarizing the nuclear spins can prolong the coherence times of electron spins. 
Such a polarization of nuclear spins is achieved by transferring spin from the electron spins to the nuclear spins in a procedure called dynamical nuclear polarization (DNP) [\onlinecite{Rudner4}]. 

A suitable system for conducting DNP is a gate defined double quantum dot loaded with two electrons.
There has been a variety of proposals [\onlinecite{Levy}, \onlinecite{b2}] to use DQDs as qubits, e.g., by focusing on the singlet $|S\rangle=1/\sqrt{2}(|\uparrow\rangle|\downarrow\rangle$ $-|\downarrow\rangle|\uparrow\rangle)$ and triplet $|T_0\rangle=1/\sqrt{2}(|\uparrow\rangle|\downarrow\rangle+|\downarrow\rangle|\uparrow\rangle)$ 
logical subspace [\onlinecite{r16}], where the generated nuclear difference field and the exchange interaction are used to perform universal control of the qubit on the Bloch sphere. Other than the already mentioned DNP, the effects of dephasing caused 
by a nuclear spin bath, can be canceled by applying a Hahn echo sequence [\onlinecite{Hahn}], or the more elaborate CPMG sequences [\onlinecite{r16}]. 

The generation of a nuclear gradient field, required to control the $S-T_0$ qubit [\onlinecite{r16}], can be achieved by cycling the electron spins over the anticrossing between the singlet $|S\rangle=1/\sqrt{2}(|\uparrow\rangle|\downarrow\rangle-|\downarrow\rangle|\uparrow\rangle)$ and triplet $|T_+\rangle=|\uparrow\rangle|\uparrow\rangle$ states. 
During such a $S-T_+$ cycle, the electron spins transfer polarization to the nuclear spins [\onlinecite{g7}], and a nuclear difference field is generated.
Furthermore, a higher degree of nuclear spin polarization causes a longer spin coherence time of the $S-T_0$ qubit. In materials with sizable spin-orbit interaction, the spin-orbit interaction induces electron spin flips, and this mechanism competes with the hyperfine mediated electron spin flips required for DNP.
In such materials, we theoretically explore the interplay of spin-orbit and hyperfine effects on nuclear spin preparation schemes, in the vicinity of the $S-T_+$ anticrossing. 

We assume that the dots are embedded in the semiconductor material In$_{x}$Ga$_{1-x}$As with $0\le x\le1 $. 
We model 150 nuclear spins per dot fully quantum mechanically, keeping track of how the probabilities and coherences of all nuclear states change in time. As compared to our model, recent models treating more [\onlinecite{g7}] or fewer [\onlinecite{f6}] nuclear spins fully quantum mechanically,  
do not take into account the spin-orbit interaction.
Although there has been some work on the interplay of spin-orbit and nuclear effects in GaAs double quantum dots [\onlinecite{c3}-\onlinecite{d4}], to our best knowledge none of these theoretical frameworks treat the nuclear spin dynamics fully quantum mechanically, nor investigate the nuclear spin dynamics when subjected 
to a large number $(\approx 300)$ of DNP cycles.
On the other hand, again to our best knowledge, there has been no theoretical work to describe the $S-T_+$ DNP in materials having strong spin-orbit interaction, e.g., InAs. Experiments in InAs have been carried out with a single electron spin in a single quantum dot [\onlinecite{d42}], 
or in a double quantum dot, by using a different, more elaborate pulsing sequence [\onlinecite{d43}].
As a consequence of our fully quantum treatment we can give precise estimations of $T_{2}^{*}$, compare them to known experiments in GaAs [\onlinecite{e5}], and calculate a value for $T_{2}^{*}$ in In$_{x}$Ga$_{1-x}$As. 
Our results can also be be extrapolated to materials with even stronger spin-orbit coupling as compared to InAs such as, e.g., InSb. 

This paper is organized as follows. In Section \ref{secmod} we describe our model, in Section \ref{secdic} we discuss the total nuclear spin angular momentum basis which significantly reduces the dimension of our Hilbert space. In Section \ref{secevo} 
we study the time evolution during the DNP cycle, in Section \ref{results1} we present results on In$_{0.2}$Ga$_{0.8}$As, a material with an intermediate strength of spin-orbit interaction, and in Section \ref{results2} 
we compare results for different abundances of indium in In$_{x}$Ga$_{1-x}$As. We conclude in Section \ref{seccon}.

\section{MODEL}\label{secmod}
\begin{figure}[t!]
\centering
\includegraphics[height=6.0cm]{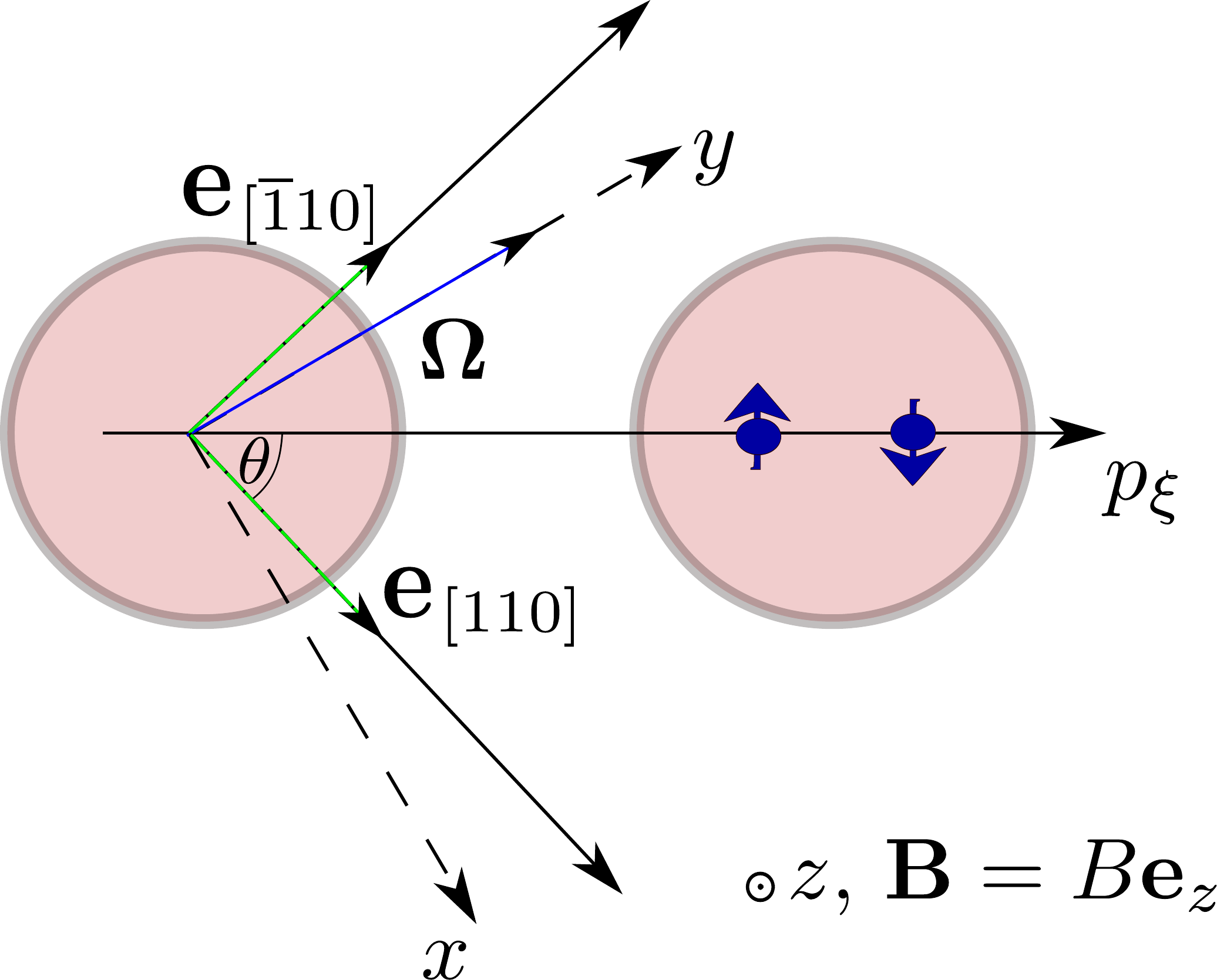}
\caption{(Color online) Geometry of the problem. The strength of spin-orbit interaction is tuned by varying the angle $\theta$ between the $[110]$ crystallographic axis and the interdot connection axis $p_{\xi}$. Spin-orbit interaction 
generates an effective magnetic field ${\bf \Omega}$ along the $y$ axis. The external magnetic field is perpendicular to the $[110]$ - $p_{\xi}$ plane.}
\label{geom}
\end{figure}

\begin{figure}[t]
\includegraphics[height=6.9cm]{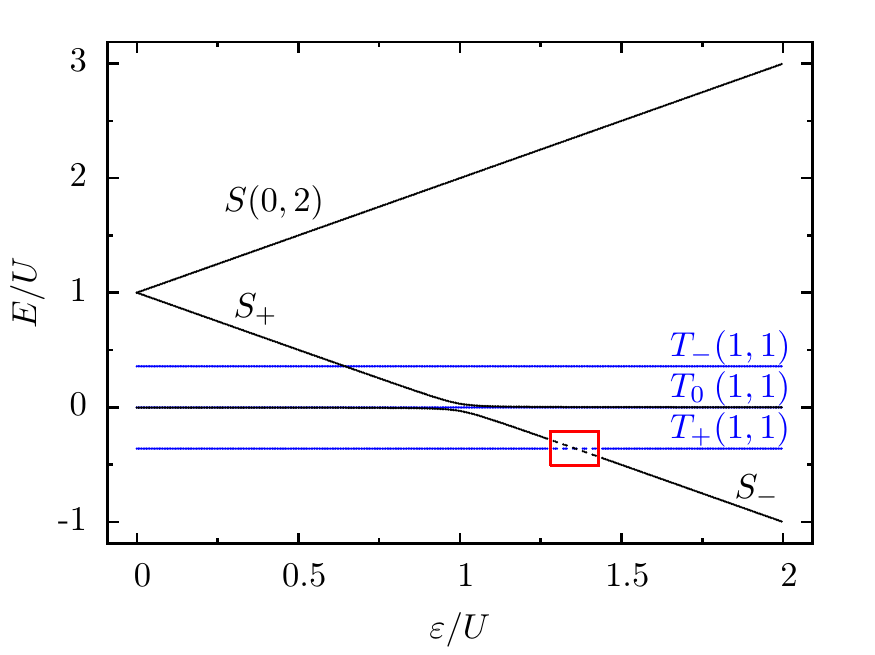}
\put(-130,0){\raisebox{3.55cm}{\mbox{\includegraphics[height=2.8cm]{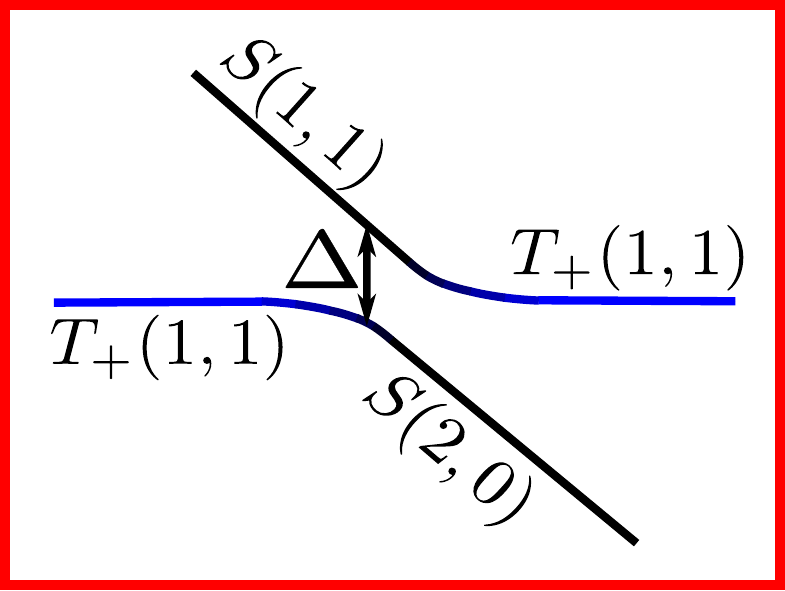}}}}
\caption{(Color online) Two-electron spectrum of a DQD in InAs as a function of the interdot bias $\varepsilon$, obtained by diagonalizing the Hamiltonian $H_{0}$ [Eq. (\ref{eq:H0})]. The energy $E$ and the detuning $\varepsilon$ are expressed in units of the Coulomb energy $U$. 
The parameters of the plot are the magnetic field $B=1\text{ T}$, the Coulomb energy $U=4.86\text{ meV}$, the extended tunneling hopping $t_H=0.11\text{ meV}$, the triplet matrix element $V_{+}=2.16\,\mu\text{eV}$, the doubly occupied singlet matrix element $V_-=0.42\,\mu{eV}$, 
half of the interdot separation $a=73.6\text{ nm}$. Including hyperfine interaction and/or spin-orbit interaction opens up an avoided crossing $\Delta$ [\onlinecite{h8}] (upper inset). The magnetic field is chosen large, as compared to the value in the remainder of the paper, for visualization purposes.
The $S(2,0)$ and $S(0,2)$ are singly occupied singlets, $S(1,1)$ is the doubly occupied singlet. $T_+$, $T_0$ and $T_-$ are triplet states corresponding to $m_s=1$, $m_s=0$ and $m_s=-1$. The $S_-$ and $S_+$ are the lower and the upper hybridized singlet 
[see Eq. (\ref{eq:eigenvectors}) and Eq. (\ref{eq:eigenvectors2})].}
\label{spectrum}
\end{figure}

The confinement in a quantum dot is modeled with a quadratic potential and the electronic wave functions are calculated according to the Hund-Mulliken theory [\onlinecite{j10}].
Our approach is a good approximation in the regime where half of the interdot separation $a$ is larger than the effective 
Bohr radius, $a\gtrsim a_{B}=\sqrt{\hbar/m^{*}\omega_{0}}$. Here, $\omega_{0}$ is the circular frequency of the confining potential, 
which we later assume to be $\hbar\omega_0=3.0 \text{ meV}$, and $m^*$ is the effective electron mass ($m^*=0.067m_0$ for GaAs and $m^*=0.023m_0$ for InAs). 
The interdot separation $2a$ needs to be chosen sufficiently large, due to the fact that the Hund-Mulliken theory is valid in the regime of weakly interacting quantum dots. 
On the other hand, the extended tunneling matrix element $t_H$ needs to be nonvanishing, so that our DNP sequence is still possible. Therefore, for In$_{0.2}$Ga$_{0.8}$As, which is the material we study in Section \ref{results1}, we want $t_H\approx 0.01U$, 
where $U$ is the Coulomb energy of the electrons. This is why we set $a=46.3 \text{ nm}$. A magnetic field of $B=110\text{ mT}$ is applied perpendicular to the plane spanned by the $[110]$ and $[\bar{1}10]$ crystallographic axes, see Fig. \ref{geom}. 
The specific value of the magnetic field is chosen so that the $S-T_+$ anticrossing is located at $\varepsilon\approx3U/2$, where $\varepsilon$ is the energy difference between the quantum dots, Fig. \ref{spectrum}.

All stable isotopes of gallium and arsenide have a nuclear spin $j=3/2$, while stable isotopes of indium have a nuclear spin $j=9/2$. Here we discuss a simplified model in which all of the nuclear spins are assumed to be $j=1/2$ [\onlinecite{o14}]. 
Also, spin-orbit effects depend strongly on the homogeneity of the distribution of In and Ga atoms in In$_x$Ga$_{1-x}$As. Here, we assume a completely homogenous distribution of In and Ga. 
For numerical convenience we model a geometry in which the $[110]$, $[\bar{1}10]$ crystallographic axes and the interdot connection axis $p_\xi$ lie in plane (Fig. \ref{geom}). 
We develop a numerical method for modeling up to $N=150$ nuclear spins per dot, a constraint imposed by our current computational resources.

The total Hamiltonian describing the electronic and nuclear degrees of freedom is
\begin{equation}
H=H_{0}(\varepsilon)+H_{\rm HF}+H_{\rm SO}.
\end{equation} Here $H_{0}(\varepsilon)$ is the non-relativistic Hamiltonian of two electrons in a QD [\onlinecite{j10}],
\begin{widetext}

\begin{equation}\label{eq:H0}
H_{0}(\varepsilon)=
\begin{pmatrix}
U-\varepsilon & X 		& -\sqrt{2}t_H & 0 & 0 & 0\\
X	      &U+\varepsilon	& -\sqrt{2}t_H & 0 & 0 & 0\\
-\sqrt{2}t_H	      &-\sqrt{2}t_H		&V_+& 0 & 0 & 0\\
 0 	      & 0               & 0 &V_-+g\mu_BB_z\\
  0 	      & 0               & 0 &0  &V_-&0\\
  0&	0&0&0&0&V_--g\mu_BB_z
\end{pmatrix},
\end{equation}
\end{widetext} in the basis of $\{S(2,0),S(0,2),S(1,1),T_+(1,1), T_0(1,1),\\T_-(1,1)\}$. The letter $S$ denotes the singlet state, and $T_+$, $T_-$, $T_0$ are triplet states with the total spin projections $m_s=+1$, $m_s=-1$, $m_s=0$. The numbers in 
the parentheses indicate the charge state. More specifically, $(2,0)$ denotes a state where the left dot is occupied with two electrons and the right dot is empty, $(0,2)$ denotes a state where the right dot is being occupied with two electrons and the left dot is empty, 
and $(1,1)$ stands for each dot being occupied with one electron. The Hamiltonian [Eq. (\ref{eq:H0})] acquires time dependence through the bias energy $\varepsilon$. To describe the DNP process, the bias energy $\varepsilon$ will be assumed to be a linear function 
of time $\varepsilon=r  t,$ where we set $r=2U/\tau$, and where $\tau=50\text{ ns}$ is the duration of the bias sweep. The value of $r$ is chosen so that $\varepsilon=2U$ at the beginning of the sweep ($t=0$), $\varepsilon=0$ at the and of the sweep ($t=\tau$), 
as in the experiment by Petta {\it et al.} [\onlinecite{1}]. 
  
The quantities in $H_0$ are the on-site Coulomb energy $U\sim1\text{ meV}$, the coordinated hopping from one dot to the other $X\sim0.1\text{ $\mu$eV}$, the doubly occupied singlet and triplet matrix elements, $V_+,\,V_-\sim10\text{ $\mu$eV}$, 
and the extended hopping parameter, $t_H \sim0.01 U$ [\onlinecite{j10}]. The Zeeman energy is given as $g \mu_B B_z$, where $g$
is the electron $g$ factor ($g=-0.44$ for GaAs, $g=-14.7$ for InAs), the Bohr magneton is $\mu_B=5.79\times10^{-5}\text{ eV/T}$ and $B_z=110\text{ mT}$ is the magnetic field. For an electron confined in an GaAs QD the Zeeman energy at this field is $E_z=2.8\times10^{-6}\text{ eV}$.
Due to the fact that we are interested in the $S-T_+$ transition, we focus our attention on the energy subspace spanned by the states $\{S(2,0)$, $S(1,1)$, $T_+(1,1)\}$. The singlet $S(0,2)$ is high in energy with respect to the other two singlets [cf. Fig. \ref{spectrum}]
(for positive values of the detuning $\varepsilon$) whereas the remaining two singlets $S(2,0)$ and $S(1,1)$ are close in energy. The triplet states $T_{0}(1,1)$, and $T_-(1,1)$ are split off from the $T_+(1,1)$ by the Zeeman energy.
It should be mentioned that we treat the Hamiltonian [Eq. (\ref{eq:H0})] using the adiabatic approximation, meaning that the system will remain in its instantaneous eigenstates. This allows us to obtain the eigenenergies by diagonalizing the Hamiltonian
$H_0$ in the subspace of $\{S(2,0),S(1,1)\}$.
As a result of the diagonalization we obtain the two hybridized singlets $|S_{+}\rangle$, $|S_{-}\rangle$ [\onlinecite{j10}, \onlinecite{h8}] with energies
\begin{equation}
E_{S_{\pm}}=\frac{U-\varepsilon+V_{+}}{2}\pm\sqrt{\frac{(U-\varepsilon+V_{+})^2}{4}+2t_H^2},
\end{equation}
and eigenvectors
\begin{equation} \label{eq:eigenvectors}
|S_-\rangle=c(\varepsilon)|S(1,1)\rangle+\sqrt{1-c(\varepsilon)^2}|S(2,0)\rangle,
\end{equation}
\begin{equation}\label{eq:eigenvectors2}
|S_+\rangle=\sqrt{1-c(\varepsilon)^2}|S(1,1)\rangle-c(\varepsilon)|S(2,0)\rangle.
\end{equation}
With $c(\varepsilon)=\cos{\psi}$ we denote the charge admixture coefficient which can be expressed with the charge admixture angle $\psi$, where
\begin{equation}
\cos{2\psi}=\frac{U-V_{+}-\varepsilon}{\sqrt{(U-V_{+}-\varepsilon)^2+8t_H^2}}. 
\end{equation}
We only take into account the transitions between the lower hybridized singlet $|S_{-}\rangle$ and triplet $|T_+\rangle$ because the upper hybridized singlet $|S_+\rangle$ is higher in energy, and therefore can be neglected, as shown in Fig. \ref{spectrum} . 

The spin-orbit term $H_{\rm SO}$ in the Hamiltonian is a function of the angle $\theta$ [cf. Fig \ref{geom}] between the $[110]$ crystallographic axis and the interdot connection axis $p_{\xi}$ [\onlinecite{h8}],  
\begin{equation}\label{eq:SO}
H_{\rm SO}=\frac{i}{2}{\bf \Omega}(\theta)\cdot\sum\limits_{s,t=\uparrow,\downarrow}(c_{Ls}^{\dagger}\boldsymbol\sigma^{st}c_{Rt}-h.c.),
\end{equation}\\
 where ${\bf \Omega}(\theta)$ 
is the spin-orbit effective magnetic field defined by
\begin{equation}
i{\bf \Omega}(\theta)=
\langle\Phi_{L}|\hat{p}_{\xi}|\Phi_{R}\rangle((\beta-\alpha)\cos\theta{\bf e}_{[\bar{1}10]}+(\beta+\alpha)\sin\theta{\bf e}_{[110]}).
\end{equation}\\
\noindent{
Here $\alpha$ and $\beta$ are the Rashba [\onlinecite{y23}] and Dresselhaus [\onlinecite{z24}] coefficients,
the $c_{r,s}^{\dagger}$ operator creates an electron with spin $s=\uparrow,\downarrow$, in the right or left dot, $r=R,L$. Further, $\boldsymbol\sigma^{s,t}$ is the 
vector of Pauli matrices and $\Phi_{L,R}$ are the spatial parts of the wavefunctions corresponding to the left and the right dot respectively [\onlinecite{h8}] and $\hat{p}_{\xi}$ is the component of the momentum operator along the interdot connection axis. 

For computational simplicity, we choose our coordinate system such that the matrix elements of the spin-orbit part of the Hamiltonian [Eq. (\ref{eq:SO})] are always real. This is achieved by setting the ${\bf e}_y$ axis of our coordinate
system parallel with ${\bf \Omega}$ [\onlinecite{h8}], as shown in Fig. \ref{geom}. When the spin-orbit interaction is excluded, our $x$ and $y$ axes are parallel to the crystallographic axes.

Finally, the hyperfine part of the Hamiltonian is given by [\onlinecite{g7}]
\begin{equation}
 H_{\rm HF}={\bf S_{1}\cdot h_{1}+S_{2}\cdot h_{2}}=\frac{1}{2}\sum\limits_{i=1}^{2}(2S_i^zh_i^z+S_i^+h_i^-+S_i^-h_i^+),
\end{equation}
where $S_i^{(\pm)}$ are the $i$th electron spin ladder operators, $S_i^z$ and $h_i^z$ are the $z$ components of the $i$th electron 
spin operator and Overhauser field operator. Furthermore, $h_{i}^{\pm}=h_i^x\pm ih_i^y$ are the ladder operators of the Overhauser field,
\begin{equation}
{\bf h}_{i}=\sum\limits_{k=1}^{n(i)}A_{i}^{k}{\bf I}_{i}^{k},
\end{equation} where ${\bf I}_{i}^{k}$ are the nuclear spin operators for the $k$th nuclear spin in contact with the $i$th electron spin.
The strength of the hyperfine coupling between the $i$th electron and the $k$th nuclear spin is labeled $A_i^{k}$. In general $A_i^{k}$ can have a different 
value for every nuclear spin, but we simplify this by assuming a constant hyperfine coupling $A_i^{k}=A^{\rm tot}/N$ [\onlinecite{f6}].

Performing a diagonalization in the singlet subspace spanned by $\{S(2,0)$, $S(1,1)\}$, we find that the singlet eigenfunctions are bias dependent and therefore time dependent [Eq. (\ref{eq:eigenvectors}) and Eq. (\ref{eq:eigenvectors2})]. This implies that the coupling between the lower hybridized singlet $|S_{-}\rangle$ and the $|T_+\rangle$ triplet state is time dependent 
as compared to time independent coupling between the $|S(1,1)\rangle$ and $|S(2,0)\rangle$ singlets and the $|T_+\rangle$ triplet. The time dependence of the coupling
originates on the fact that the coupling depends on the charge state of the hybridized singlet [Eq. (\ref{eq:eigenvectors}) and Eq. (\ref{eq:eigenvectors2})]. The state $S(2,0)$ couples to $T_+$ only via the spin-orbit interaction and $S(1,1)$ couples to $T_+$ only by means of the hyperfine interaction. 
By using the wavefunctions of the lower hybridized singlet (see Eq. (\ref{eq:eigenvectors}) we can calculate the matrix element of the Hamiltonian between the lower hybridized singlet $|S_{-}\rangle$ and the triplet $|T_+\rangle$
\begin{equation}
\begin{split}
\langle S_{-}|H |T_+\rangle=&c(\varepsilon)\langle S(1,1)|H_{\rm HF}|T_+\rangle\\
                           &+\sqrt{1-c(\varepsilon)^2}\langle S(2,0)|H_{\rm SO}|T_+\rangle.
\end{split}
\end{equation}

It should be mentioned that due to time dependent interactions, the model discussed here must go beyond the Landau-Zener model [\onlinecite{k11}-\onlinecite{m13}].

\section{THE BASIS OF TOTAL ANGULAR MOMENTUM}\label{secdic}

In our model, all nuclear spins are treated as having spin $j=1/2$. This means that the total number of nuclear spin states is ${\rm dim}(\mathcal{H})=2^{N}$, where $N$ is the number of
nuclear spins in a quantum dot. Because the total number of nuclear spin states scales exponentially with $N$ it would be impossible to treat a large number ($N=150$) of nuclear spins with the computational power at our disposal. 
In order to make the problem treatable we first make a basis change from the product basis$\{\uparrow,\text{ }\downarrow\}$, to the basis of total angular momentum $\{|j,m\rangle\}$. 
Here $j$ is the total nuclear spin quantum number, $0 \le j \le N/2$, and $m$ is the total nuclear spin projection along the $z$ axis, $-j\le m \le j$. 
Now the total number of states can be written as 
\begin{equation}\label{eq:perm}
{\rm dim}(\mathcal{H})=\sum\limits_{j=0}^{N/2}\sum\limits_{\rm perm}(2j+1)=2^N. 
\end{equation} The inner sum runs over all permutation symmetries for a given value of $j$. 
The basis of total angular momentum still scales as ${\rm dim}(\mathcal{H})=2^N$, but now certain states in the inner sum in Eq. (\ref{eq:perm}) do not need to be taken into account, and states with higher 
$j$ in the outer sum in Eq. (\ref{eq:perm}) can be neglected due to
the low probability of their occurrence. In the remainder of this
section we will describe in more detail how we reduce the number of
nuclear spin states from 
${\rm dim}(\mathcal{H})=2^N$ 
to ${\rm dim}(\mathcal{H}^\prime)\ll 2^N$.

Neither the hyperfine nor the spin-orbit interaction mix states with different $j$, and thus the matrix representing our Hamiltonian is block diagonal with every block corresponding to a value of $j=j_0,\,j_0+1,\,\ldots N/2$. The value of $j_0$ depends 
on the parity of $N$, for an even $N$, $j_0=0$ and for an odd $N$, $j_0=1/2$. The probability distribution of nuclear spin states, with respect to the quantum number $j$ is a Gaussian (in the limit $N\rightarrow \infty$) with its maximum located at 
$\approx \sqrt{N/2}$, Fig. \ref{gauss}. From now on we will refer to this value of $j$ as its most likely value, $j_{\rm ml}\approx\sqrt{N/2}$. The nuclear spin probability distribution, with respect to the number of nuclear spins per dot $N$ and quantum number 
$j$ is given by the following formula [\onlinecite{m13b}]
\begin{center}
\begin{equation}\label{eq:statedistribution}
 p(N,j)= \frac{(2j+1)^{2}N!}{(N/2+j+1)!(N/2-j)!2^{N}}.
 \end{equation}
 \end{center}

The $j$ and $m$ quantum numbers are generally not sufficient to describe all possible nuclear spin states. Other than $j$ and $m$, the nuclear spin states are described by their permutation symmetries. 
For example, for three nuclear spins defined by quantum numbers $j=1/2$ and $m=1/2$, there are two states $|1/2,1/2\rangle$ and $|1/2,1/2\rangle^\prime$ with distinct permutation symmetries. These two states are not mixed by homogenous hyperfine or by spin-orbit interactions.
Furthermore, they remain equally probable as the matrix elements of the Hamiltonian only depend on $j$ and $m$ and not on the symmetry properties. 
Therefore, by evaluating our system for a certain symmetry $|1/2,1/2\rangle$ we would also know the behavior of the state with a different permutation symmetry $|1/2,1/2\rangle^\prime$.
By generalizing this simple example to $N$-spin systems we can significantly reduce the number of the states we consider. For every value of $j$ we need to evaluate only one state of symmetry in Eq. (\ref{eq:perm}), 
and therefore for each value of $j$ the inner sum in Eq. (\ref{eq:perm}) can be replaced by one representing term.

\begin{figure}[t!]
\centering
\includegraphics[height=6.5cm]{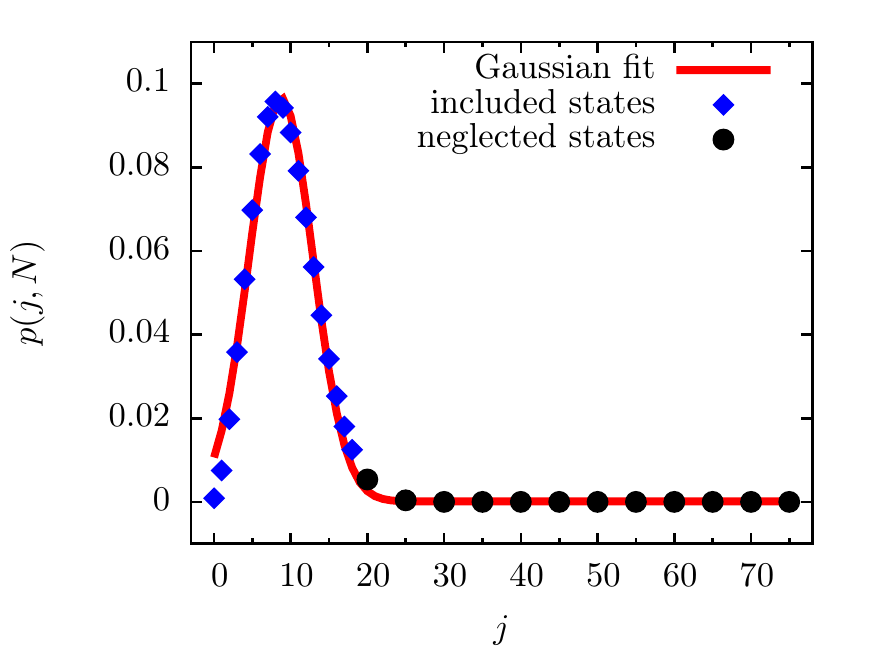}
\caption{(Color online) Initial nuclear spin probability distribution with respect to the quantum number $j$ for $N=150$ nuclear spins $1/2$, where $j_{\rm ml}=\sqrt{N/2}$ and $j_{\rm max}=18$. 
Throughout our calculations we only consider the states $0\le j \le j_{\rm max}$ (blue diamonds) and do not consider the states $j>j_{\rm max}$ (black circles).}
\label{gauss}
\end{figure}
We can reduce the number of states further by choosing the maximum value of $j$ we take into consideration, $j_{\rm max}$ in a manner that $\sqrt{N/2} \ll j_{\rm max} \ll N/2$. The omission of all states with $j>j_{\rm max}$ 
is justified because these states occur with a very low probability (see Fig. \ref{gauss} and Eq. (\ref{eq:statedistribution})). Now the total number of the states we consider scales with $j_{\rm max}$ as
\begin{equation}\label{eq:perm2}
{\rm dim}(\mathcal{H}^\prime)=\sum\limits_{j=0}^{j_{\rm max}}(2j+1)\cong(j_{\rm max}+1)^2\ll2^N. 
\end{equation}
Due to the fact that the states with different $j$ do not mix by any interaction we consider, we can analyze our system for one value of $j$ at a time and finally average over all included values of $j$.
By doing so, we average over close to (but not exactly) $100 \%$ of all possible states. In our case, $N=150$ nuclear spins per dot and $0 \le j \le 75$. Constraining ourselves to $0 \le j \le j_{\rm max}=18$, we average over $97.8\%$ of all possible nuclear spin configurations, as shown in Fig. \ref{gauss}. 
The efficiency of our approach can be illustrated best if we calculate the number of states in the $\{\uparrow$, $\downarrow\}$ basis and in the $\{|j,m\rangle\}$ basis after we consider only one symmetry state for every $j$ and consider only $0\le j\le j_{\rm max}$. 
For $N=150$, Eq. (\ref{eq:perm}) yields ${\rm dim}(\mathcal{H})\approx
1.4\times 10^{45}$ and for $j_{\rm max}=18$, Eq. (\ref{eq:perm2}) yields ${\rm dim}(\mathcal{H}^\prime)=361$.

\section{TIME EVOLUTION DURING DNP}\label{secevo}
We now describe a single step in the DNP procedure. The system is initialized in a singlet state $S(2,0)$, where both electrons are occupying the same dot. Afterwards, the electronic system is driven with a finite velocity through the $S-T_+$ anticrossing 
(see Fig. \ref{spectrum}) by varying the voltage bias $\varepsilon$. The electronic state is then measured, and finally the system is reset quickly to the initial state $S(2,0)$ [\onlinecite{g7}]. 
Accordingly, we propagate the density matrix of the system $\rho$ according to the update rule
\begin{equation}
\rho^{(i+1)}=M_{S}U\rho^{(i)}U^{\dagger}M_{S}+M_{T}U\rho^{(i)}U^{\dagger}M_{T}.
\end{equation}
Here $\rho^{(i)}$ and $\rho^{(i+1)}$ are the total density matrices before and after the $i$-th DNP step, $U$ is the unitary time evolution operator and $M_{S}$ and $M_{T}$ are the singlet and triplet projection operators [\onlinecite{2}]. 
They satisfy the relations $M_{S}+M_{T}=I,$ and $M_{S} M_{T}=0.$

After the evolution of the system, a measurement of the electronic state takes place. This measurement procedure has two outcomes: either a singlet $S$ or a triplet $T_+$ is detected. The nuclear density matrix is updated accordingly, 
\begin{equation}
\rho_{n}=P_{S}\rho_{n}^{S}+P_{T}\rho_{n}^{T},
\end{equation}
where $\rho_n$ is the nuclear density matrix and $P_{S}={\rm Tr}[M_{S} U \rho^{(i)} U^{\dagger}M_{S}]$ and $P_{T}={\rm Tr}[M_{T} U \rho^{(i)} U^{\dagger}M_{T}]$ are the singlet and the triplet outcome probabilities. 
The superscripts $S$ and $T$ stand for a nuclear density matrix related to the singlet and the triplet measurement outcome. 
For a certain value of $j$ we calculate the singlet return probability $ P_{S}$, and the standard deviation of the nuclear difference field, $\sigma^{(z)}=\sqrt{\langle(\delta h^{z})^{2}\rangle-\langle\delta h^{z}\rangle^{2}}$ [\onlinecite{i9}]. 
After averaging over all included $j$, we use the standard deviation of the nuclear difference field to evaluate the $S-T_0$ spin qubit decoherence time, $T_{2}^{*}=\hbar/\sigma^{(z)}$ [\onlinecite{i9}].\\

We compute the propagator $U$ by discretizing the time interval $(0,\tau)$. 
Our model describes the passage through the anticrossing with $q=100$ equally spaced, step-like time increments.
The procedure of computing the propagator is the following: For every discrete point in time 
$t_{i}$ we compute the Hamiltonian $H(t_{i})$. We approximate the propagator for the fixed time point $t_{i}$,
\begin{equation}
U_{t_{i}}=e^{-iH(t_{i})\Delta t/\hbar},
\end{equation}
with $\Delta t=\tau/q$. By repeating the procedure for every discrete step we obtain the total time evolution operator
\begin{equation} \label{eq:prop}
U=U_{t_{q}}U_{t_{q-1}}\ldots U_{t_{1}}.
\end{equation}
\begin{figure}[t!]
\centering
\includegraphics[height=16.5cm]{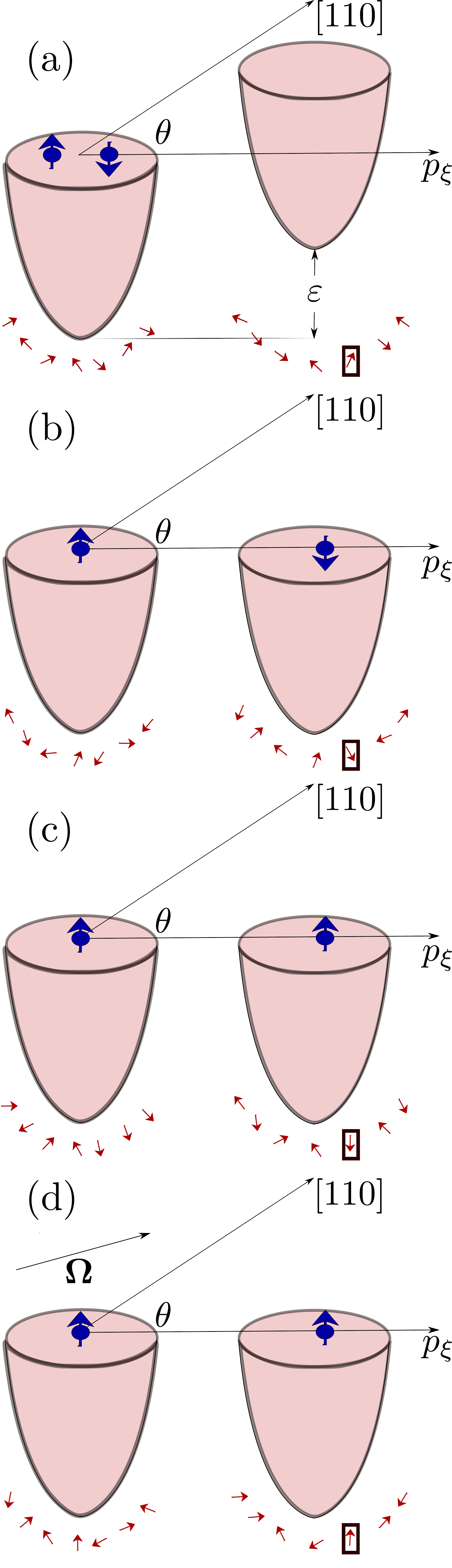}
\caption{(Color online)
System initialization and measurement outcomes. (a) Initially, the quantum dots have an energy bias $\varepsilon$ and the two electrons rest in a singlet $(2,0)$ state on the left dot. 
(b) After slowly tuning $\varepsilon$ to zero, and measuring a singlet outcome, due to the weak measurement the spin of the nuclear bath decreases.
(c) In the case of a spin triplet outcome an electron spin flips and the spin of the nuclear bath is changed accordingly.
(d) The electronic spin can also be flipped due to spin-orbit, and the spin of the nuclear bath is pumped in the opposing direction (up) due to the weak measurement. With $\varepsilon$ we denote the voltage bias, 
$\theta$ is the angle between the $[110]$ crystallographic axis and the interdot connection axis $p_{\xi}$, ${\bf \Omega}$ is the spin-orbit effective magnetic field.}
\label{viz}
\end{figure}
Tuning the system across the $S-T_+$ point and measuring the electronic state
after every forward sweep changes the probabilities and coherences of the electronic and the nuclear states.
The qualitative picture is simpler if we first disregard the spin-orbit interaction. When the spin-orbit interaction is excluded, both the electronic spin singlet and the triplet outcomes
 increase the probability for nuclear spins to be in the spin down state [\onlinecite{g7}], corresponding to generating negative values of nuclear spin polarization $P=(n_{\uparrow}-n_{\downarrow})/(n_{\uparrow}+n_{\downarrow})$, where $P$ is the nuclear spin polarization, 
$n_{\uparrow}$ is the number of nuclear spins pointing up and $n_{\downarrow}$ is the number of nuclear spins pointing down [cf. Figs. \ref{viz}(a-d)]. 

There is one more possible process, involving spin-orbit interaction, which is not shown in Fig. \ref{viz}. After cycling the electronic system across the $S-T_+$ anticrossing the system can end up in a virtual $T_+$ state due to spin-orbit interaction, 
but is instantaneously transferred to a singlet state due to hyperfine interaction, 
accompanied by a flip of the nuclear spin from down to up, thus changing the nuclear spin polarization closer to positive values. This is a process that, along with the process visualized on Fig. \ref{viz}(d), competes with the hyperfine-mediated 
generation of negative polarization of the nuclear spins (down pumping). These two processes combined compensate the down pumping in systems with strong spin-orbit interaction.  

To make an effective comparison between In$_{x}$Ga$_{1-x}$As systems with different indium content $x$ we keep the same values for $B_z$ and $d=a/a_B=2.186$. This
implies that the single particle tunneling and the overlap between the quantum dots would remain the same for every value of $x$ (see Ref. [\onlinecite{j10}]). For a comparison between different materials, the relative strength of the spin-orbit interaction 
can be quantified by the ratio of $\Xi=4a/\Lambda_{\rm SO}$, where $\Lambda_{\rm SO}$ is the spin-orbit length defined by 
\begin{equation}\label{eq:LambdaSO}
\frac{1}{\Lambda_{\rm SO}}=\frac{m^*}{\hbar}\sqrt{\cos^2{\theta}(\alpha-\beta)^2+\sin^2{\theta}(\alpha+\beta)^2}.
\end{equation} Here, $m^*$ is the effective electron mass, $\alpha$ and $\beta$ are the Rashba and Dresselhaus constants and $\theta$ is the angle between the $[110]$ crystallographic axis and the interdot connection axis $p_{\xi}$ [cf. Fig.\ref{viz}].

The spin-orbit length is the distance which an electron needs to travel in order to have its spin flipped due to spin-orbit interaction. If the electrons are initialized in a singlet state the probability for flipping the tunneling electron 
due to spin-orbit interaction is $P_{\rm flip}=1/2$ at $2a=\Lambda_{\rm SO}/2$. This further implies that if $\Xi<1$, the system is more probable to remain in a singlet state. If $\Xi=1$ the $S$ and $T_+$ outcomes due to 
spin-orbit coupling are equally probable and finally if $1<\Xi<2$ a $T_+$ outcome due to spin-orbit is more probable, because the probability that the tunneling electron has flipped its spin is greater than  $P_{\rm flip}>0.5$. 
In our study $\Lambda_{\rm SO}/2\gg2a$ which implies $\Xi\ll1$, thus
singlet outcomes due to spin-orbit interaction are always more probable even in pure InAs with the
strongest possible value of spin-orbit ($\theta=\pi/2$). In pure InAs,
with $\theta=\pi/2$, $\Xi\approx0.63$ for $d=a/a_B=2.186$. 
 
\section{RESULTS FOR I\lowercase{n}$_{0.2}$G\lowercase{a}$_{0.8}$A\lowercase{s}}\label{results1}
\begin{figure}[t!]
\setlength{\belowcaptionskip}{5pt}
\centering
\hspace*{-0.15cm}\includegraphics[height=4.25cm]{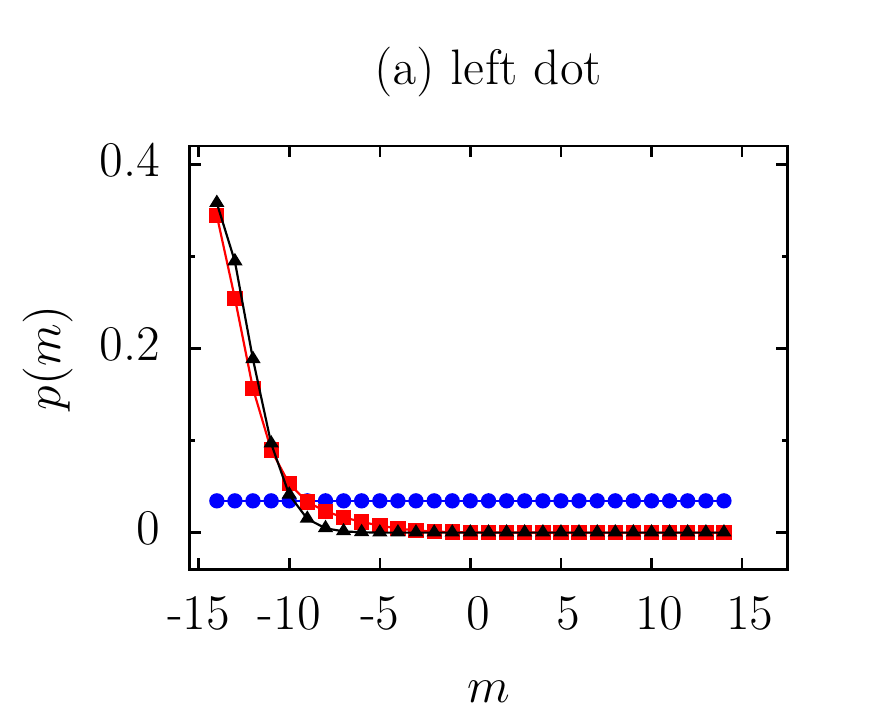}%
\hspace*{-1.11cm}\includegraphics[height=4.25cm]{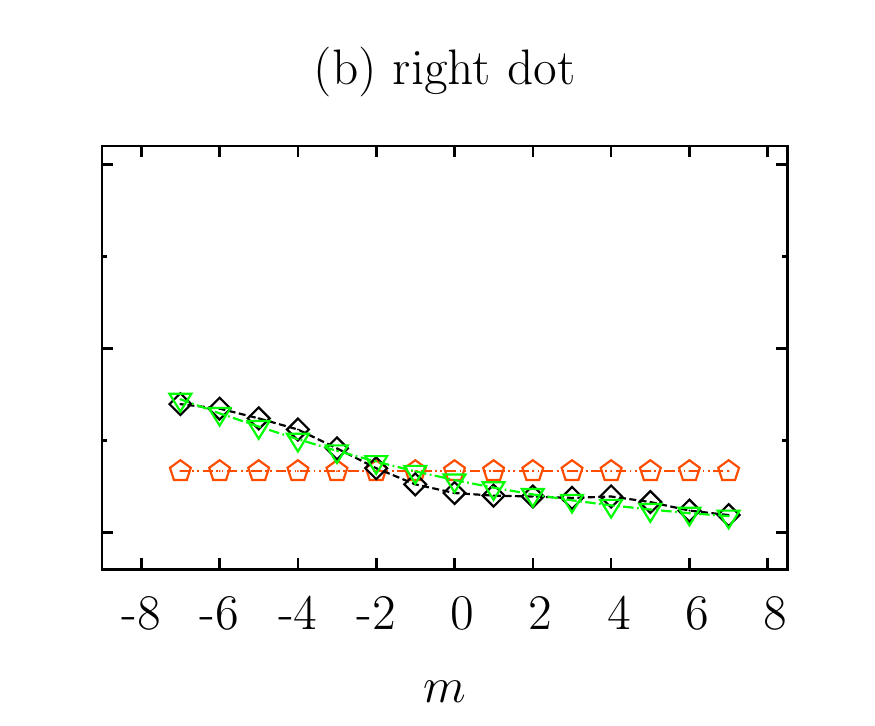}%
\caption{ (Color online) (a) Probability distribution in the left quantum dot with respect to the nuclear spin projection quantum number $m$ for $j_{L}=14$. Blue circles represent the initial probability distribution, black triangles represent
the probability distribution after 300 cycles with spin-orbit interaction excluded, and red squares represent the probability distribution after 300 cycles with spin-orbit interaction included. (b) Probability distribution in the right 
quantum dot with respect to the nuclear spin projection quantum number $m$ for $j_{R}=7$. Red pentagons present the initial probability distribution, green triangles represent the probability distribution after 300 cycles with spin-orbit interaction excluded 
and black diamonds represent the probability distribution after 300 cycles when spin-orbit interaction is included corresponding to $\theta=\pi/2$. Here, $\theta$ is the angle between the 
$[110]$ crystallographic axis and the interdot connection axis $p_{\xi}$. The number of nuclear spins per quantum dot is $N=150$.}
\label{DistL-R}
\end{figure}
Our attention is now focused on In$_{0.2}$Ga$_{0.8}$As, a material with an intermediate strength of spin-orbit coupling, as compared to the relatively weak spin-orbit coupling in GaAs and relatively strong spin-orbit coupling in InAs. 
We have evaluated the system of $N=150$ nuclear spins per dot, for different values of the angle $\theta$ and with $j_{\rm max}=18$.
 States with $j>j_{\rm max}$ would further lower the $T_2^*$ and $P_s$ and increase $\sigma^{(z)}$.
Therefore, we point out that our results provide an upper bound for $T_2^*$ (including states with $j>j_{max}=18$ could lower $T_2^*$ for at most $2.2\%$, see Fig. \ref{gauss} and Eq. (\ref{eq:statedistribution})) and $P_s$ and a lower bound for $\sigma^{(z)}$. 
We study the effect of 300 DNP cycles on the nuclear spin state. We find that the spin-orbit interaction has a notable effect on nuclear state preparation. In Fig. \ref{DistL-R}, we plot the probabilities of nuclear spin states for a case with a given value of 
$j^{L,R}$ in the left and the right dot.

For $j^{L}=14$ and $j^{R}=7$ the pumping procedure has altered the nuclear probability distribution from a uniform distribution (with respect to the quantum number
$m$) to a probability distribution where states with negative $m$ are more likely. In the case without spin-orbit interaction, two processes contribute to this negative pumping of the nuclear spin [\onlinecite{g7}] - 
the singlet detection accompanied by a weak measurement of the nuclear spin state and the $T_+$ detection, which flips the nuclear spin down to conserve the total spin of the system [cf. Fig. \ref{viz}(b) and Fig. \ref{viz}(c)]. 
Although including spin-orbit interaction [cf. Fig. \ref{DistL-R}(a), Fig. \ref{DistL-R}(b)], changes the final distribution of nuclear spin states only slightly, spin-orbit effects still have a 
notable effect on the singlet return probability $P_{S}={\rm Tr}[M_{S} U \rho ^{(i)}U^{\dagger}M_{S}]$. In Fig. \ref{Ps}, we plot $P_S$ as a function of the number of cycles across the $S-T_+$ anticrossing for In$_{0.2}$Ga$_{0.8}$As. 
Here, we tune the strength of the spin-orbit interaction by varying the angle $\theta$ between the $[110]$ crystallographic axis and the interdot connection axis $p_{\xi}$.
\begin{figure}[t!]
    \centering
    \includegraphics[height=6.5cm]{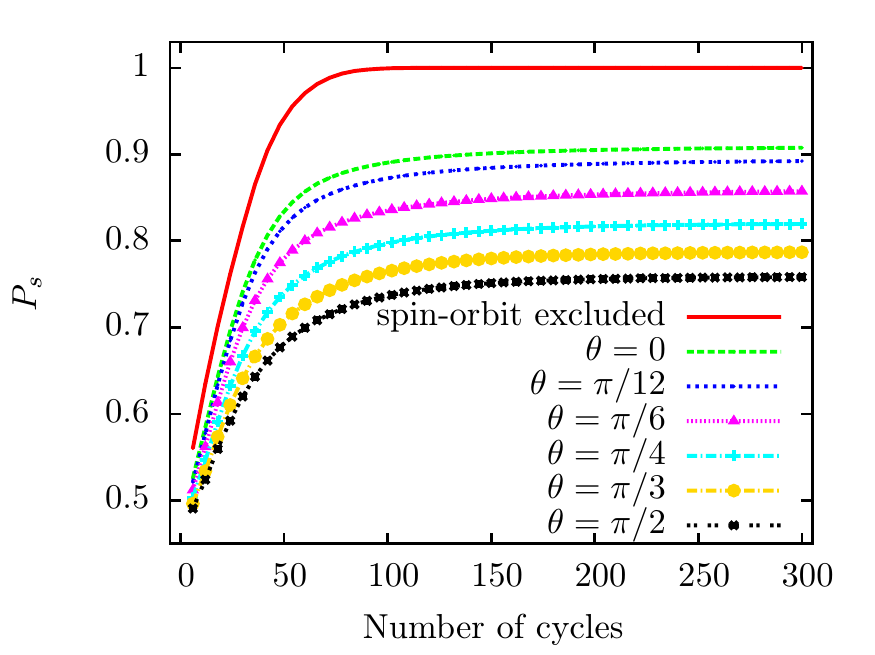}
    \caption{(Color online) The singlet return probability $P_S$ as a function of the number of cycles across the $S-T_+$ anticrossing in In$_{0.2}$Ga$_{0.8}$As. Here, $\theta$ is the angle between the $[110]$ crystallographic axis and the interdot connection 
    axis $p_{\xi}$.}
    \label{Ps}
\end{figure}
As shown in Fig. \ref{Ps} (solid red line), repeatedly cycling the system across the anticrossing point polarizes the nuclear spins, which leads to $P_{s}=1$ after 300 cycles [\onlinecite{g7}]. The situation
changes dramatically when we include the spin-orbit interaction, which competes with the hyperfine mediated down pumping of the nuclear spin.

By theoretically varying the strength of the spin-orbit interaction, we find that 
when the spin-orbit interaction has the largest possible value for $\theta=\pi/2$, it significantly affects the singlet return probability $P_{s}\approx0.72$ (Fig. \ref{Ps}).
Including spin-orbit interaction generates a mechanism which polarizes nuclear spins in the up direction (see Section \ref{secevo} and Fig. \ref{DistL-R}).
As a consequence of this behavior, the nuclear preparation mechanism is not efficient when spin-orbit effects are strong. The interplay of the hyperfine and spin-orbit interactions on nuclear state preparation can be observed better if we plot the standard deviation of the 
nuclear difference field $\sigma^{(z)}$ (Fig. \ref{sigmaz}). We notice that the spin-orbit interaction has prevented the reduction of the standard deviation of the nuclear difference field ($0 \le \theta \le\pi/2$, see Fig. \ref{sigmaz}).
\begin{figure}[t!]
\includegraphics[height=6.5cm]{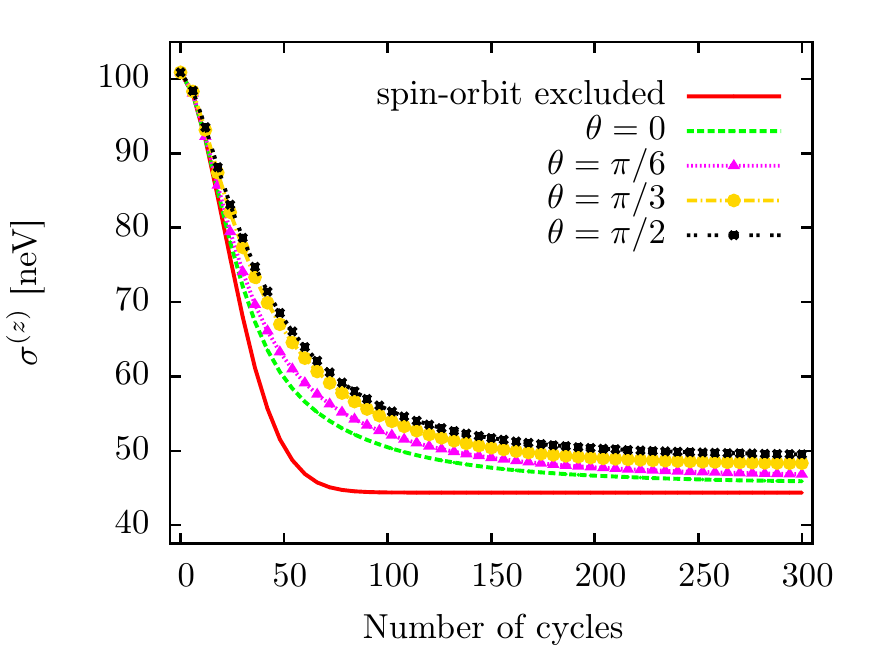}
\caption{(Color online) Standard deviation of the nuclear difference field $\sigma^{(z)}$ with respect to the number of DNP cycles across the $S-T_+$ anticrossing and different values of angle $\theta$ in In$_{0.2}$Ga$_{0.8}$As. Here, $\theta$ is the angle between the $[110]$ crystallographic axis and the interdot connection axis $p_{\xi}$.}
\label{sigmaz}
\end{figure}
Spin-orbit interactions affect the efforts to increase the spin $S-T_0$ qubit decoherence time $T_{2}^{*}$, see Fig. \ref{T2*}. The strongest spin-orbit coupling, corresponding to $\theta=\pi/2$,
slightly lowers the resulting decoherence time from $T_2^*\approx 15 \text{ ns}$ (red line) to $T_2^*\approx 13 \text{ ns}$ (black dashed line with black $x$ symbols).

Without the spin-orbit interaction our theory predicts that the ratio of the final decoherence time (after the cycling is complete) $T_{2,f}^{*}$ and initial decoherence time (before the cycling starts) $T_{2,i}^{*}$ is  $T_{2,f}^{*} / T_{2,i}^{*}\approx2.28$ [cf. Fig. \ref{10^6}]. 
The situation changes when we include spin-orbit interaction. For $\theta=0$ we find a value of $T_{2,f}^{*}/T_{2,i}^{*}\approx2.20$, while for $\theta=\pi/2$ the ratio is $T_{2,f}^{*} / T_{2,i}^{*}\approx2.04$. 

After the inclusion of the spin-orbit interaction the ratio $T_{2,f}^{*} / T_{2,i}^{*}$ decreases with $\theta$. Our results suggest that the $S-T_+$ 
dynamical nuclear polarization is not as effective in materials with intermediate strength of spin-orbit interaction, as compared to those without spin-orbit coupling. Nevertheless, the DNP still provides a notable enhancement of the $S-T_0$ qubit decoherence time $T_2^*$. 
\begin{figure}[t!]
\includegraphics[height=6.5cm]{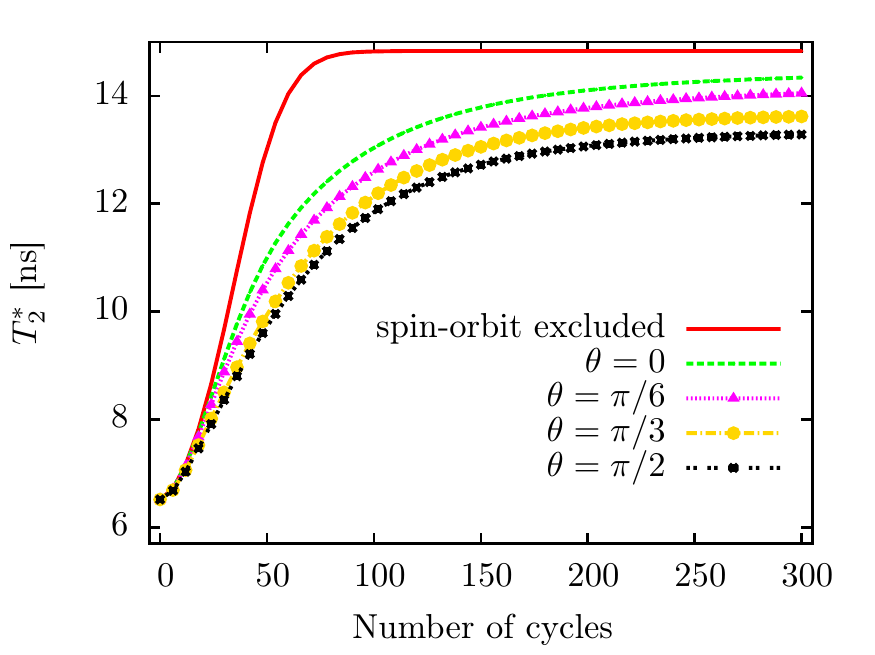}
\caption{(Color online) $S-T_0$ qubit decoherence time $T_{2}^{*}$ as a function of the number of DNP cycles across the $S-T_+$ anticrossing and strength of spin-orbit interaction in In$_{0.2}$Ga$_{0.8}$As. Here, $\theta$ is the angle between the $[110]$ crystallographic axis and the interdot connection axis $p_{\xi}$.}
\label{T2*}
\end{figure}
\begin{figure}[t!]
\centering
\includegraphics[height=6.5cm]{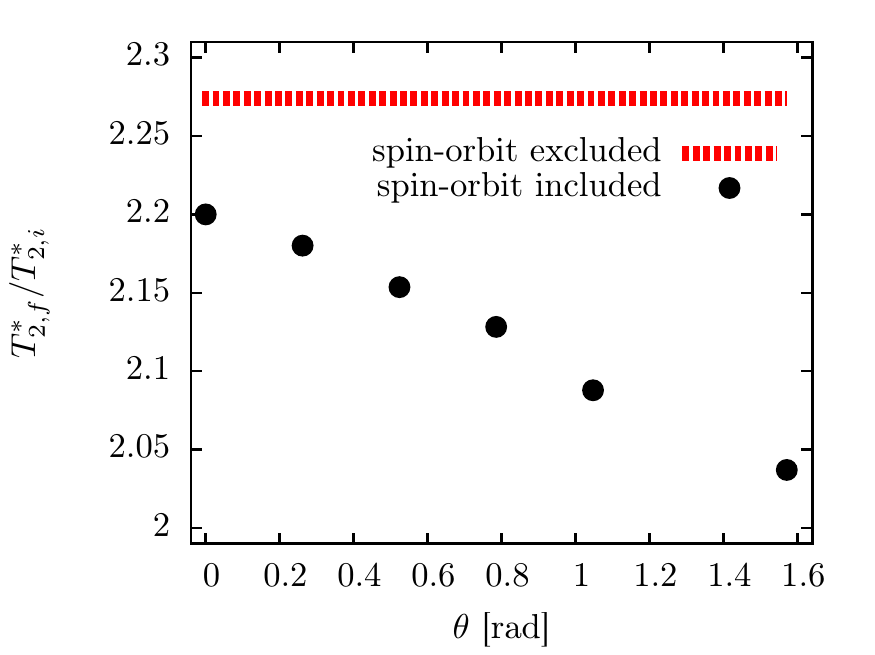}
\caption{(Color online)  The ratio of the final $T_{2,f}^*$ and initial $T_{2,i}^*$ decoherence times in In$_{0.2}$Ga$_{0.8}$As, for different values of the angle $\theta$ between the $[110]$ crystallographic axis and the interdot connection axis $p_{\xi}$.}
\label{10^6}
\end{figure}
We work in the so called "giant spin model" and we model the behavior of $10^{4}-10^{6}$ nuclear spins with significantly fewer spins, $\sim10^{2}-10^{3}$.
In general $\sigma_{i}^{(z)}\propto A_i^{k}$, which would give rise to a much higher standard deviation of the nuclear difference field than expected. Therefore, we rescale the hyperfine constant,
such that $\sigma_{i}^{(z)}$ has the same value for $N\approx10^{6}$, and $N=150$, $j_{\rm ml}=\sqrt{N/2}$. The predicted decoherence time before the start of the DNP is $T_2^{*}\approx6.2\text{ ns}$ while measurements yield $T_2^{*}\approx10\text{ ns}$ 
for pure GaAs [\onlinecite{1}] (where excluding spin-orbit effects is a good approximation). Since $\sigma_{i}^{(z)} \propto \sqrt{N}$, and $\sigma_{f}^{(z)}$ does not depend on $N$ but on different parameters, we can estimate that 
$T_{2,f}^{*} / T_{2,i}^{*} \sim \sqrt{N}$ for our case of $N=150$ and the realistic case $N=10^{6}$ (for an electrically defined quantum dot in In$_x$Ga$_{1-x}$As).
Therefore, we can estimate the maximum possible ratio of initial and final decoherence times for the realistic case of $N=10^{6}$ spins and spin-orbit interaction excluded and included to be
$T_{2,f}^{*} / T_{2,i}^{*} \approx 175$ without spin-orbit interaction, compared to $T_{2,f}^{*} / T_{2,i}^{*} \approx 94$ for GaAs in reference [\onlinecite{g7}], $T_{2,f}^{*} / T_{2,i}^{*} \approx 174$ 
for $\theta=0$, $T_{2,f}^{*} / T_{2,i}^{*} \approx 163$ for
$\theta=\pi/2$. 
\section{RESULTS FOR I\lowercase{n}$_{x}$G\lowercase{a}$_{1-x}$A\lowercase{s}} \label{results2}

In this section we will compare the $T_2^*$ results for In$_{x}$Ga$_{1-x}$As with varying In content $x$. We vary the concentration of indium $x$ in the sample between $0$ and $1$ with a $0.2$ increment. For the sake of computational efficiency, and the fact that we are interested in a mere comparison between materials with different percentages of indium,  our computational method is 
slightly simplified now. Instead of averaging over all possible states ranging from $j_{\rm min}-j_{\rm max}$ we set $j^L=j^R=j_{\rm ml}=\sqrt{N/2}$ for the left and the right quantum dot. 
This effectively means that we are simulating a situation where an experiment is performed only once with the most likely nuclear spin configuration.

\begin{figure}[t!]
\centering
\includegraphics[height=6.5cm]{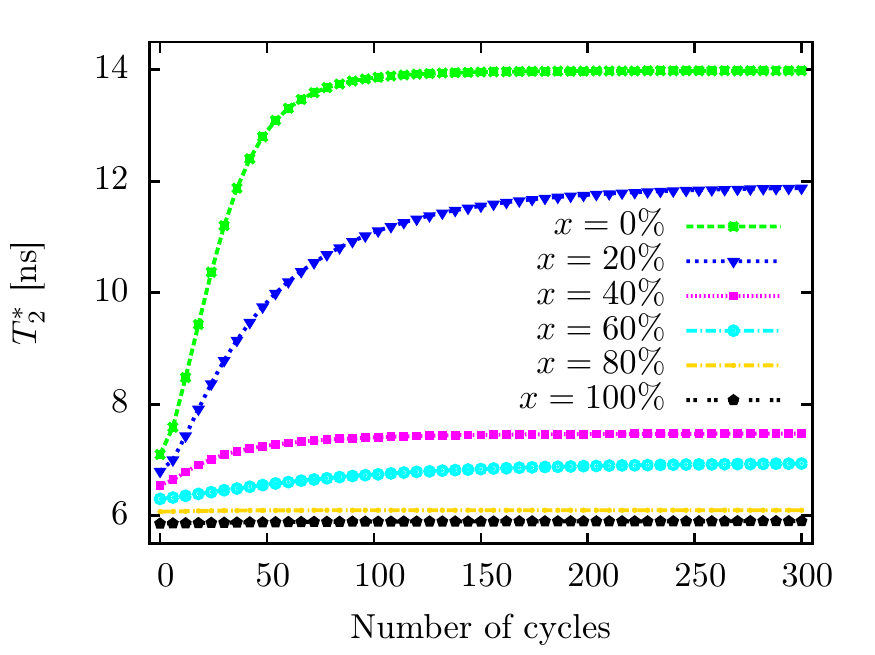}
\caption{(Color online)  $S-T_0$ electron spin coherence time $T_2^*$ as a function of the number of DNP cycles across the $S-T_+$ anticrossing, for different abundances of indium $x$ in In$_{x}$Ga$_{1-x}$As and for $\theta=\pi/2$. Here, $\theta$ is the angle between the $[110]$ crystallographic axis and the interdot connection axis $p_{\xi}$. }
\label{InAsAll}
\end{figure}

From Fig. \ref{InAsAll} we conclude that raising the concentration of indium in a In$_{x}$Ga$_{1-x}$As sample has a detrimental effect on the efficiency of the $S-T_+$ DNP scheme. By doping the system with indium, the Rashba spin-orbit coupling
is strengthened, thus reducing the overall $\Lambda_{\rm SO}$ [Eq. (\ref{eq:LambdaSO})], which as a consequence has more virtual and real $T_+$ outcomes due to the spin-orbit interaction. The virtual $T_+$ will relax to $S$, quickly flipping a nuclear spin from down to up in the process. The real spin-orbit 
mediated $T_+$ outcomes will also pump the nuclear spin towards the positive values of the polarization (up). This process can completely vain efforts to increase $T_2^*$, 
even at intermediate concentrations of 40$\%$ In
(Fig. \ref{InAsAll}). At higher indium concentrations, DNP is totally
suppressed for all values of $\theta$ [cf. Fig. \ref{GaAsvsInAsT0}].

\begin{figure}[t!]
\centering
\includegraphics[height=6.5cm]{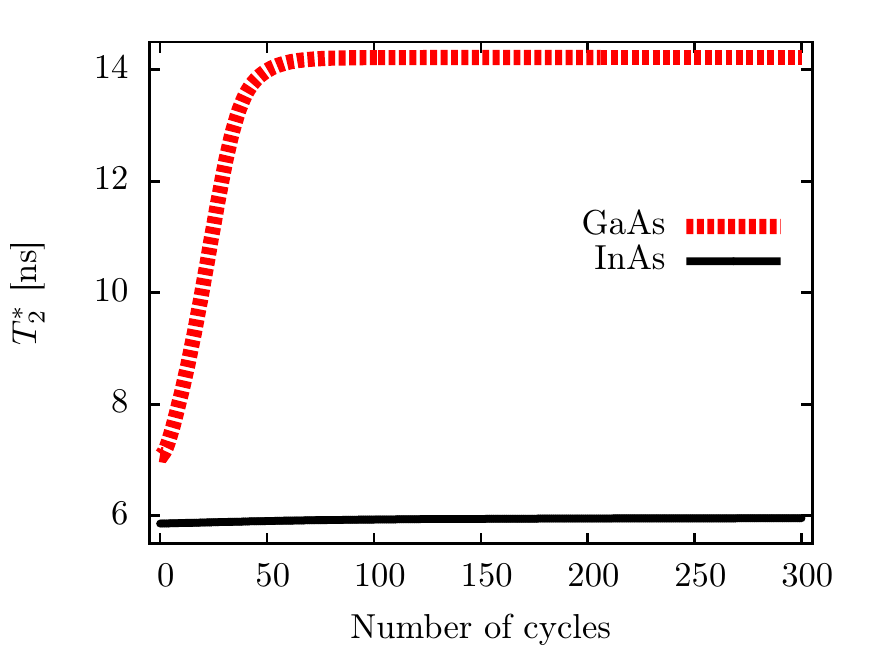}
\caption{(Color online)  $S-T_0$ electron spin coherence time $T_2^*$ for GaAs and InAs as a function of the number of DNP cycles across the $S-T_+$ anticrossing, for $\theta=0$, i.e. the case where the $[110]$ crystallographic axis and the interdot connection axis $p_{\xi}$ are aligned.}
\label{GaAsvsInAsT0}
\end{figure}
\section{CONCLUSIONS AND FINAL REMARKS}\label{seccon}
Our results show that pure InAs is a not a suitable candidate for $S-T_+$ DNP, due to the fact that the enhancement of $T_2^*$ is strongly suppressed even for the smallest possible strength of the spin-orbit interaction corresponding to $\theta=0$.
Dynamical nuclear polarization in InAs could still be achieved by using single spin single quantum dot systems [\onlinecite{d42}] or by using a more elaborate pulsing sequence [\onlinecite{d43}]. A similar behavior could be expected in materials with even stronger spin-orbit as
compared to InAs and that is, e.g., InSb.

To conclude, we have discussed a nuclear polarization scheme in In$_{x}$Ga$_{1-x}$As double quantum dots with spin-orbit interaction included. In the presence of spin-orbit interaction a suppression of the enhancement of $T_{2}^{*}$ is predicted. 
Our conclusions are also valid for materials with fewer nuclear spins. We underline that the $S-T_+$ DNP sequence is highly sensitive to the strength of the spin-orbit coupling, and therefore the efficiency of the $S-T_+$ DNP sequence will depend on the angle $\theta$ 
and the In content $x$ in In$_x$Ga$_{1-x}$As. A stronger spin-orbit interaction will establish a process that will quickly neutralize any efforts to prolong $T_2^*$. The cases of unequally coupled and/or sized dots, and different shapes of the bias [\onlinecite{r16}] 
are in general treatable by our numerics and will be the subject of our future studies. Charge noise [\onlinecite{p15}-\onlinecite{s17}] is neglected in the current model. Investigating the significance of charge coherence requires an extension of the numerical tools
we use [\onlinecite{q16}], and is planned as a forthcoming investigation.

\begin{acknowledgments}
We thank Hugo Ribeiro for useful discussions and the EU S$^3$NANO Marie Curie ITN and Deutsche Forschungsgemeinschaft (DFG) within the SFB 767 for financial support.
\end{acknowledgments}


\end{document}